\documentstyle[preprint,aps]{revtex}

\begin{document}
\title{Paramagnetic limiting of the upper critical field of the layered organic
superconductor $\kappa $--(BEDT-TTF)$_{2}$Cu(SCN)$_{2}$}
\author{F.~Zuo$^{1}$, J. S. Brooks$^{2}$, Ross H. McKenzie$^{3}$, J.~A.~Schlueter$%
^{4}$, and Jack~M.~Williams$^{4}$}
\address{$^{1}$Department of Physics, University of Miami, Coral Gables, Florida
33124 }
\address{$^{2}$National High Magnetic Field Laboratory, Tallahassee, Florida 32306}
\address{$^{3}$School of Physics, University of New South Wales, Sydney 2052,
Australia}
\address{$^{4}$Chemistry and Materials Science Divisions,\\
Argonne National Laboratory, Argonne, Illinois 60439}
\maketitle

\begin{abstract}
We report detailed measurements of the interlayer magnetoresistance of the
layered organic superconductor $\kappa $--(BEDT-TTF)$_{2}$Cu(SCN)$_{2}$ for
temperatures down to 0.5 K and fields up to 30 tesla. The upper critical
field is determined from the resistive transition for a wide range of
temperatures and field directions. For magnetic fields parallel to the
layers, the upper critical field increases approximately linearly with
decreasing temperature. The upper critical field at low temperatures is
compared to the Pauli paramagnetic limit, at which singlet superconductivity
should be destroyed by the Zeeman splitting of the electron spins. The
measured value is comparable to a value for the paramagnetic limit
calculated from thermodynamic quantities but exceeds the limit calculated
from BCS theory. The angular dependence of the upper critical field shows a
cusp-like feature for fields close to the layers, consistent with decoupled
layers. \newline
\end{abstract}


\section{Introduction}

The layered organic molecular crystals $\kappa $-(BEDT-TTF)$_{2}$X (where
BEDT-TTF is bis-(ethylenedithia-tetrathiafulvalene) and X is an anion (e.g.,
X=I$_{3}$, Cu[N(CN)$_{2}$]Br, Cu(SCN)$_{2}$)) are particularly interesting
because they are strongly correlated electron systems with similarities to
the high-$T_{c}$ cuprate superconductors including unconventional metallic
properties and competition between antiferromagnetism and superconductivity 
\cite{mck1,mck2,fuku,kanoda0}. Furthermore, they are available in high
purity single crystals and, in contrast to the cuprates, their lower
superconducting transition temperature ($T_{c}\sim 10$K) makes
experimentally accessible in steady magnetic fields properties such as the
upper critical field and Shubnikov-de Haas oscillations\cite{ish,wos}.

Recently it has been argued that a minimal theoretical model that can
describe these materials is a Hubbard model on an anisotropic triangular
lattice with one hole per site\cite{mck2,fuku}. Calculations at the level of
the random-phase approximation\cite{rpa} and the fluctuation-exchange
approximation\cite{flex} suggest that at the boundary of the
antiferromagnetic phase this model exhibits superconductivity mediated by
spin fluctuations. As the anisotropy of the intersite hopping varies the
model changes from the square lattice to the isotropic triangular lattice to
decoupled chains\cite{mck2}. The wavevector associated with the
antiferromagnetic spin fluctuations changes\cite{merino} and the
superconductivity has been predicted to change from d-wave singlet (as in
the cuprates) to s-wave triplet in the odd-frequency channel\cite{rpa}.

Experimental results that are consistent with unconventional
superconductivity include the temperature dependence of the nmr relaxation
rate $1/T_{1}$ (including the absence of a Hebel-Slichter peak)\cite
{desoto,kanoda}, the temperature and magnetic field dependence of the
electronic specific heat\cite{nakazawa}, the temperature dependence of the
thermal conductivity,\cite{belin} and the sensitivity of $T_{c}$ to disorder 
\cite{zuo}.

The temperature dependence of the nmr Knight shift (which measures the
electron spin susceptibility) in the superconducting state provides a means
to distinguish triplet and singlet pairing. For triplet pairing the Knight
shift does not change on entering the superconducting state, whereas for
singlet pairing the Knight shift goes to zero as the temperature decreases
to zero. The Knight shift of $^{13}$C nmr on the X=Cu[N(CN)$_2$]Br is
consistent with the latter. In contrast, the Knight shift of $^{17}$O nmr on
Sr$_2$RuO$_4$ is consistent with the former.\cite{ishida}

If the superconductivity is spin singlet then the upper critical field
cannot exceed the paramagnetic limit $H_{P}$, also known as the Pauli limit
or Clogston-Shandrasekhar limit\cite{clogston,pauli}. Above $H_{P}$ the
Cooper pairs are destroyed by the Zeeman splitting produced by the magnetic
field coupling to the electronic spins. For weak-coupling BCS theory 
\begin{equation}
H_{P} = H_{P}^{BCS} \simeq {\frac{1.8k_{B}T_{c} }{\mu _{B}}}.  \label{bpauli}
\end{equation}
For $T_c = 10$ K, as in the material studied here, this gives $H_{P}^{BCS} =
18 $ T. Strong coupling effects\cite{perez} and d-wave pairing\cite{sondhi}
only change this value of $H_{P}$ slightly. In most superconductors the
paramagnetic limit is irrelevant because the superconductivity is destroyed
at much lower fields due to the frustration of the orbital degrees of
freedom associated with the formation of vortices. However, in layered
superconductors with fields parallel to the layers the vortices can fit
between the layers and paramagnetic limiting can become important.\cite
{klemm}

Previous determinations of the upper critical field of the $\kappa $%
-(BEDT-TTF)$_{2}$X family\cite{oshima,kwok,lang,graebner,murata,delong} have
mostly focussed on measurements of the slope ${\frac{dH_{c_{2}}(T)}{dT}} $
near $T_{c}$. The values obtained for X=Cu[N(CN)$_{2}$]Br and X=Cu(SCN)$_{2}$
are in the range 10 to 20 T/K. Using the WHH formula\cite{whh} for a
three-dimensional superconductor, this very large slope would suggest a
zero-temperature $H_{c_{2}}(T=0)=0.7T_{c}{\frac{dH_{c_{2}}(T)}{dT}=}$ 70-140
T, which is well above the BCS Pauli limit. A previous transport measurement
on the X=Cu(SCN)$_{2}$ salt was carried out in pulsed magnet fields.\cite
{saito} A quasi-linear temperature dependence was found with $H_{c2}\sim $25
T and the authors concluded that the upper critical field exceeded the Pauli
limit. A study of the upper critical field of X=Cu(CN)[N(CN)$_2$]\cite
{nakamura} determined from the resistive transition found an upper critical
field of about 25 T for fields parallel to the layers. Studies on the lower T%
$_{c}$ organic compounds such as the $\kappa $-(BEDT-TTF)$_{2}$I$_{3}$\cite
{wanka}, $\beta $-(BEDT-TTF)$_{2}$I$_{3}$, and $\beta $-(BEDT-TTF)$_{2}$IBr$%
_{2}$\cite{murata2} have found that the $H_{c2}$ at zero temperature lies
below or close to the Pauli paramagnetic limit predicted by BCS theory.
Similar paramagnetic field limited $H_{c2}$ have been reported in the
cuprate YBa$_{2}$Cu$_{3}$O$_{7-\delta }$\cite{dzurak} and the heavy fermion
superconductors UPd$_{2}$Al$_{3}$\cite{gloos}.

If there is paramagnetic limiting there is theoretically the possibility
that as the magnetic field is increased at low temperatures there is a
first-order phase transition into non-uniform superconducting state,
originally proposed by Fulde and Ferrell and Larkin and Ovchinikov\cite{ff}.
As the dimensionality of the system decreases the magnetic field range over
which this phase is stable increases.\cite{shimahara} Such a first-order
phase transition was recently seen in ultrathin beryllium films.\cite{adams}
It is still controversial about whether this phase does exist in UPd$_{2}$Al$%
_{3}$\cite{gloos}. On the other hand, if the superconductivity is triplet
there is also the possibility of re-entrant superconductivity at high fields
such that $T_{c}(H)$ actually increases with increasing field\cite
{dupuis,lebed2}.

In this paper we report the measurement of the interlayer resistivity of
X=Cu(SCN)$_{2}$ down to 0.5 K and up to 30 T for a range of field
directions. For magnetic fields parallel to the layers, the upper critical
field increases approximately linearly with decreasing temperature to values
that clearly exceed the BCS Pauli limiting field (\ref{bpauli}), but are
consistent with the paramagnetic limit, estimated directly from the
superconducting condensation energy. The upper critical field as a function
of angle shows a sharp cusp for fields almost parallel to the layers,
consistent with two-dimensional decoupled layers. We find no evidence of a
first order phase transition as a function of field at low temperatures.

\section{Theoretical background}

We now briefly summaries some theoretical results concerning the upper
critical field which we will use later in interpreting our results. A more
complete discussion can be found in Reference \onlinecite{bul}.

\subsection{Angular dependence of the upper critical field}

Anisotropic Ginzburg-Landau theory is valid when the coherence length
perpendicular to the layers, $\xi _{\perp }$, is much larger than the
interlayer spacing. It predicts that the dependence of the upper critical
field on the angle $\theta $ between the field and the normal to the layers
is\cite{klemm,bul,schneider} 
\begin{equation}
\left[ \frac{H_{c2}(\theta )\cos (\theta )}{H_{c2\perp }}\right] ^{2}+\left[ 
\frac{H_{c2}(\theta )\sin (\theta )}{H_{c2\parallel }}\right] ^{2}=1,
\label{anisot}
\end{equation}
where $H_{c2\perp }$ and $H_{c2\parallel }$ are the upper critical field for
fields perpendicular and parallel to the layers, respectively. The
perpendicular upper critical field is determined by $\xi _{\parallel }$, the
coherence length parallel to the layers, 
\begin{equation}
H_{c2\perp }={\frac{\Phi _{0}}{2\pi \xi _{\parallel }^{2}}}  \label{hcperp}
\end{equation}
where $\Phi _{0}$ is the flux quantum. The coherence lengths parallel and
perpendicular to the layers are related by 
\begin{equation}
{\frac{\xi _{\parallel }}{\xi _{\perp }}}={\frac{H_{c2\perp }}{%
H_{c2\parallel }}}.  \label{ratio}
\end{equation}

Klemm, Luther, and Beasley considered the upper critical field of layered
superconductors when the layers were infinitely thin.\cite{klemm} For both
Lawrence-Doniach theory and microscopic theory, they found that for fields
parallel to the layers, if the interlayer coupling is sufficiently weak the
upper critical field diverges at low temperatures unless spin-orbit effects
or paramagnetic limiting is present. This is because the Josephson vortices
associated with the field parallel to the layers have no normal core and can
fit between the layers. Bulaevskii\cite{bul} and Schneider and Schmidt\cite
{schneider} considered a more general model where the layers have a finite
thickness $d$, resulting in a finite upper critical field 
\begin{equation}
{\frac{H_{c2\perp }}{H_{c2\parallel }}}={\frac{d}{\sqrt{12}\xi _{\parallel }}%
}.  \label{ratio2}
\end{equation}
They also found that if the coupling between the layers is sufficiently weak
then the angular dependence of the upper critical field is given by 
\begin{equation}
\left| \frac{H_{c2}(\theta )\cos (\theta )}{H_{c2\perp }}\right| +\left[ 
\frac{H_{c2}(\theta )\sin (\theta )}{H_{c2\parallel }}\right] ^{2}=1
\label{tink}
\end{equation}
This same angular dependence was found earlier for thin two-dimensional
films by Tinkham using a simple fluxoid quantization argument.\cite{tinkham}
The main difference from the anisotropic three-dimensional result is that at 
$\theta =$90$^{\text{o}}$, $H_{c2}(\theta )$ from Eqn. (\ref{anisot}) is
smooth or bell-shaped with $\frac{dH_{c2}(\theta )}{d\theta }=0$ whereas $%
H_{c2}(\theta )$ from Eqn. (\ref{tink}) has \ a cusp at $\theta =$90$^{\text{%
o}}$.

If the upper critical field is determined solely by coupling of the field to
the spins then it will be independent of the field direction. Bulaevskii\cite
{bul} considered the case where the paramagnetic limit is larger than the
upper critical field for fields perpendicular to the layers but smaller than
the upper critical field determined by orbital effects for fields parallel
to the layers. The angular dependence is then given by 
\begin{equation}
\left| \frac{H_{c2}(\theta )\cos (\theta )}{H_{c2\perp }}\right| \left[1-
\left(\frac{H_{c2\perp}}{H_{c2\parallel }} \right)^2 \right] +\left[ \frac{%
H_{c2}(\theta )}{H_{c2\parallel }}\right] ^{2}=1  \label{pauli-tink}
\end{equation}
where $H_{c2\parallel }=H_P$. This also results in an $H_{c2}$ versus $%
\theta $ curve which has a cusp at $\theta =$90$^{\text{o}}$. Indeed the
angular dependence is difficult to distinguish from Eqn. (\ref{tink}).

\subsection{ Estimating the paramagnetic limiting field}

The metallic phase has a finite Pauli spin susceptibility $\chi _{e}$
compared to the vanishing susceptibility (at zero temperature) of a spin
singlet superconducting state. Hence, it will be energetically favorable to
destroy the superconducting state when the magnetic energy density gained by
the difference in susceptibilities exceeds the superconducting condensation
energy density $U_{c}$. The critical field $H_{P}$ at which this occurs is
given by\cite{clogston} 
\begin{equation}
U_{c}={\frac{\mu _{0}}{2}}\chi _{e}H_{P}^{2}  \label{cond}
\end{equation}
where $\mu _{0}$ is the magnetic permeability of free space.

In BCS theory the condensation energy density is $U_c = {\frac{1 }{2}}
N(E_F) \Delta(0)^2 $ where $N(E_F)$ is the metallic density of states and $%
\Delta(0)=1.76 k_B T_c$ is the zero-temperature energy gap. Making using of
these relations and $\chi_e = (\mu_B)^2 N(E_F)$ we obtain the expression (%
\ref{bpauli}) for $H_P$.

{\it Many-body effects}. In the $\kappa$-(BEDT-TTF)$_2$X crystals there are
significant many-body effects; the electron effective mass $m^*$ determined
from magnetic oscillations can be two to five times larger than that
predicted by band structure calculations.\cite{mck2,wos} The effect of this
on the paramagnetic limit needs to be taken into account. Perez-Gonzalez\cite
{perez} finds that the paramagnetic limiting field is enhanced by a factor
of $m^*/m_b$. However, he did not take into account the simultaneous effect
on the Zeeman splitting: the $g$ factor changes to $g^*$. When this is done
one finds that within a Fermi liquid framework the Pauli limit is actually
reduced from (\ref{bpauli}) by a factor of $g^*/g$.\cite{mck0} This ratio
can be estimated from thermodynamic measurements or from the spin-splitting
of magnetic oscillations.\cite{mck0} The values obtained by these two
methods for X=Cu(SCN)$_2$, are 0.8 and 1.4, respectively.\cite{mck0}

Alternatively, we can make a {\it theory-independent} estimate of $H_P$ by
using (\ref{cond}) and the experimentally determined condensation energy
density and spin susceptibility. This method of determining $H_P$ is very
attractive because it does include all the many-body effects (without
assuming a Fermi liquid picture) and does {\it not} assume the validity of
any particular theory of superconductivity for the material in question.
Haddon et al.\cite{haddon} found $\chi_e = 4.3 \times 10^{-4}$ emu per mole
(corresponding to a density of states of 7 states per (eV molecule)) for the
X=Cu(SCN)$_2$ salt. By a reanalysis of Graebner et al.'s\cite{graebner}
specific heat data Wosnitza\cite{wos} evaluated the condensation energy
density in terms of the thermodynamic critical field $B_{th} = $ 90 mT where 
$U_c = {\frac{ 1 }{2 \mu_0}} B_{th}^2$. 
Taking the unit cell volume of 1695 $\AA^3$ and two (BEDT-TTF)$_2$X units in
each unit cell 
gives $B_P = 30 \pm 5 $ T. The uncertainty is estimated based on the
uncertainty in the values for the condensation energy and the susceptibility.

\section{Experimental details}

Single crystals of $\kappa $-(BEDT-TTF)$_{2}$Cu(SCN)$_{2}$ were synthesized
by the electrocrystallization technique described elsewhere\cite{zuo}. The
interlayer resistance was measured with use of the four probe technique.
Contact of the gold wires to the sample was made with a Dupont conducting
paste or graphite paste. Typical contact resistances between the gold wire
and the sample were about 10 $\ \Omega $. A current of 1 $\mu A$ was used to
ensure linear {\it I-V} characteristics. The voltage was detected with a
lock-in amplifier at low frequencies of about 312 Hz. To avoid pressure
effects due to solidification of grease, the sample was mechanically held by
thin gold wires. The data presented in this work were taken in a $^{3}$He
system with field up to 30 T at the National High Magnetic Field Laboratory
at Tallahassee. The sample can be rotated in the field and the orientation
was determined by using a Hall probe at low fields.

\section{Results}

Shown in Fig. 1 is a typical field dependence of the interlayer resistance
plotted in a semi-log scale at a temperature of 4.2 K. The field is applied
parallel to the planes. The resistive transition in parallel field is
typical of the low dimensional organic superconductors with a broad
transition width in field and a large positive magnetoresistance in the
normal state. The superconducting transition or the upper critical field 
{\it H}$_{c2}$ is defined at the 1 $\Omega $ level. To check the validity of
this criteria, the critical field will be compared with that obtained by a
more conventional definition. Shown in the inset are the same data in a
linear scale. The two lines are extrapolations of the normal state
magnetoresistance and the superconducting transition with the upper critical
field {\it H}$_{c2}${\it *} defined at the crossing point of the two lines.

Fig. 2 is an overlay of resistive transitions in parallel field at different
temperatures from T = 0.5 K to 10.2 K. With increasing temperature, the
curves shift to the left toward lower critical fields. The transition curves
are nearly parallel for all temperatures in the semilog scale. {\it H}$_{c2}$
is almost the mid-transition point as in a conventional superconductor,
where parallel transitions are seen but in a linear scale.

The temperature dependences of the two fields {\it H}$_{c2}$ and {\it H}$%
_{c2}${\it *} are shown in Fig. 3. Within the scatter of the points, the two
upper critical fields have nearly the same linear temperature dependence
with {\it dH}$_{c2}${\it /dT} $\approx 3$ TK$^{-1}$. The offset in the
superconducting transition temperature is due to the different definitions.
The upper critical fields at zero temperature are about 30 T and 33 T for 
{\it H}$_{c2}$ and {\it H}$_{c2}${\it *,} respectively. The dashed line is
the Pauli limit $H_{\text{{\it P}}}$ = 18.4 T, calculated from Eqn. (\ref
{bpauli}) with{\it \ T}$_{c}$ = 10 K. Clearly, $H_{\text{{\it P}}}$ defined
this way is well below the measured upper critical fields at low
temperatures. On the other hand, $H_{c2}$ is consistent with our estimate of 
$H_{\text{{\it P}}}$ from thermodynamic quantities.

To look at the anisotropy of the upper critical field, a systematic
measurements have been taken as a function of angle $\theta ,$ defined
between the field direction and the normal of the plane. Plotted in Fig. 4
is an overlay of resistive transitions as a function of field at different
angles. The six curves are representative of the angular dependence from
field parallel to the layers ($\theta =90^{o})$ to normal to the layers ($%
\theta =180^{o})$. With increasing $\theta $, the field dependence of the
resistive transition is drastically changed. At $\theta $ = 91.50$^{\text{o}%
} $, {\it H}$_{c2}$ is decreased by about 4 T. At $\theta $ = 96.64$^{\text{o%
}} $, a shoulder-like feature is developed in {\it R(H)} with a
corresponding decrease in {\it H}$_{c2}$ by about 12 T. The shoulder-like
structure develops into a well defined peak at $\theta $ = 178$^{\text{o}}$
with the occurrence of the Shubnikov de-Hass (SdH) oscillation in the
resistance at high fields. It should be noted that unlike for fields
parallel to the layers, the resistive transition is relatively insensitive
to the angles near $\theta $ = 180$^{\text{o}}$.

The inset in Fig. 4 shows an expanded view of the resistive transitions at
angles close to $\theta =$90$^{\text{o}}$ direction. With a slight increment
in $\theta $, the transition is drastically broadened. 
The field component parallel to planes is almost constant for all angles
shown in the inset and the maximum out of plane field component is about 0.5
T at $\theta $ = 91.50$^{\text{o}}$ and H = 30 T.

{\it H}$_{c2}$ defined at the 1 $\Omega $ level as a function of angle is
shown in Fig. 5 at a temperature of 1.56 K. Clearly, {\it H}$_{c2}$
decreases rapidly away from the parallel to the plane direction and is
nearly saturated above 140$^{\text{o}}$. The three lines are fits to the
three-dimensional anisotropic model (Eqn. (\ref{anisot})) and the decoupled
layer results (Eqn. (\ref{tink}) and (\ref{pauli-tink})). The fits to Eqn. (%
\ref{tink}) and (\ref{pauli-tink}) are indistinguishable in the scale shown.
While all three fits seem reasonable at first sight, clear deviations are
seen very close to $\theta =$90$^{\text{o}}$, as shown in the inset. A
cusp-like feature is observed experimentally, as in the fit to the decoupled
layer model, while the 3D fit is rounded with a negative curvature at the
top. A better agreement with the data for the decoupled layer model at large
angles is also evident with the 3D fit lying systematically under the data.
At 1.56 K, the 2D fit gives $H_{c2\perp }$ = 2.27 T and $H_{c2\parallel }$
=24.5 T. Wanka et al.\cite{wanka} also found that the angular dependence for
the X=I$_{3}$ salt was fit best by Eqn. (\ref{tink}).


\section{Discussion}

Our value of $H_{c2\perp }= 2.3$ T at 1.56 K can be compared with the value
of about 1.8 T found from the irreversibility line deduced from torque
measurements.\cite{sasaki} From the perpendicular upper critical field value
of 2.3 T and Eqn. (\ref{hcperp}) we deduce an intralayer coherence length of
120 $\AA$. 
The anisotropic three-dimensional theory (Eqn. (\ref{ratio})) and the
measured ratio of the upper critical fields gives a perpendicular coherence
length of $\xi_\perp \simeq 13 \AA$. Since this is comparable to the
interlayer spacing of $15 \AA$ we cannot expect the theory to apply. Hence,
it is not surprising that the angular dependence is not described by Eqn. (%
\ref{anisot}).

If instead we consider the model of weakly coupled layers and use Eqn. (\ref
{hcperp}) for the ratio of the critical fields we deduce that the thickness
of the superconducting layer is $d = 40 \AA.$ Clearly, this is unrealistic
because it should be smaller than the interlayer spacing. A more realistic
value would be a few $\AA$. This suggests that the parallel upper critical
field being determined by paramagnetic limiting rather than orbital effects
is more realistic.

Because of the extremely sensitive angular dependence of the resistive
transition, a shoulder-like feature is developed in the resistive transition
a few degrees away from the parallel to the plane direction. The upper
critical field {\it H}$_{c2}$* can only be defined close to the planes.
While the magnitude of {\it H}$_{c2}$* is larger than {\it H}$_{c2}$, as
expected, it is difficult to distinguish the 2D and the 3D models with the
available data. {\it H}$_{c2}$* decreases quasi-linearly with angle within
the errors.


The upper critical field determined from transport measurements has been
under a lot of debate in the cuprate superconductors\cite{blumberg}. For
field perpendicular to the planes, {\it H}$_{c2}(T)$ defined at certain
fractional normal state resistance typically gives rise to a positive
curvature at low temperatures. Various mechanisms have been proposed for the
unconventional temperature dependence. However, it has been suggested that
the {\it H}$_{c2}$ thus defined corresponds to the irreversibility or vortex
melting line. For fields parallel to the layers, a vortex moving along the
plane encounters negligible pinning as there is no normal core associated
with Josephson vortices. Magnetization is practically always reversible in
this orientation. The resistive onset field is clearly well separated from
irreversibility field and reflects the true upper critical field.

In the case of Sr$_{2}$RuO$_{4}$ and the quasi-one-dimensional organic
superconductor (TMTSF)$_{2}$X where X=ClO$_{4}$ and PF$_{6}$\cite{lee}, the
upper critical field in the plane has been found to exceed the Pauli limit,
calculated from BCS theory. Combined with the strong dependence of the
transition temperature on the impurity concentration and the temperature
dependence of the Knight shift, triplet pairing or p-wave has been suggested
in these systems. However, the quasi-linear temperature dependence observed
here for both $H_{c2}$ and $H_{c2}$* is remarkably different from that of Sr$%
_{2}$RuO$_{4}$ and Bechgaard salts. For both Sr$_{2}$RuO$_{4}$ and (TMTSF)$%
_{2}$ClO$_{4}$, the $H_{c2}$ is found to saturate for $T/T_c < $ 0.2--0.4.
While for (TMTSF)$_{2}$PF$_{6}$, $H_{c2}$(T) along both {\it a} and {\it b}'
axes where X=ClO$_{4}$ displays a diverging temperature dependence near $T=$
0 K.

\section{conclusions}

In summary, for fields parallel to the layers we have observed an upper
critical field determined from resistive transition which is comparable to
the paramagnetic limit estimated from thermodynamic quantities but is
considerably larger than that calculated from BCS theory. There is no
evidence of a first order transition in the field dependence of the
resistivity, which would occur if there was a transition to a Fulde-Ferrell
phase. The observed anisotropy of the upper critical field is much less than
would be predicted by a model without paramagnetic limiting. The upper
critical field determined is quasi-linear with temperature. The angular
dependence of the resistive transition is consistent with the highly
anisotropic nature of the title compound with a cusp-like angular dependence
for field near the plane.

\acknowledgements

We thank P. Coleman, A. Dzurak, J. Merino, J. O'Brien and J. Wosnitza for
helpful discussions. This work was supported in part by NSF grant No.
DMR-9623306 and the Petroleum Research Fund ACS-PRF 33812-AC5. Work at the
National High Magnetic Field Laboratory was supported by NSF Cooperative
Agreement No. DMR-9016241 and the state of Florida. Work at UNSW was
supported by the Australian Research Council. Work performed at Argonne
National Laboratory was supported by the U.S. Department of Energy, Office
of Basic Energy Sciences, Division of Materials Sciences, under Contract No.
W-31-109-ENG-38.

\bigskip \newpage

Figure captions

\bigskip

Fig.1 Determination of the upper critical field. The main figure shows the
interlayer resistance as a function of magnetic field on a semi-logarithmic
scale, the upper critical field being defined as the field at which the
resistance is 1 ohm. In the inset the upper critical field $H_{c2}$* is
determined by linear extrapolation. The temperature is 4.2 K and the field
is parallel to the layers.

Fig.2 The field dependence of the interlayer resistance is shown for various
temperatures. The field is parallel to the layers.

Fig.3 The temperature dependence of the upper critical field for fields
parallel to the layers. The dashed line marks the Pauli paramagnetic
limiting field predicted by BCS theory. The two values of the upper critical
field correspond to the two different methods of determination (see Fig. 1).
The solid lines are guides for eyes.

Fig.4 Dependence of the resistive transition on the field direction.The
field dependence of the interlayer resistance is shown at various field
directions. The angles given denote the angle between the field and the
normal to the layers. The temperature is 1.56 K. The inset shows how the
resistive transition becomes significantly broader as the field is moved
slightly away from the plane of the layers, for which $\theta =90^{o}.$

Fig. 5 Dependence of the upper critical field on the field direction. The
dashed line is a fit to the anisotropic three-dimensional Ginzburg-Landau
result (Eqn. (\ref{anisot})), and the solid line to the results for weakly
coupled layers. (The curves corresponding to Eqn. (\ref{tink}), (\ref
{pauli-tink}) which neglect and include paramagnetic limiting, respectively
are indistinguishable.) The inset is an expanded view near $\theta =90^{o}$,
showing that the weakly coupled layer models give the best fit. The
temperature is 1.56 K.

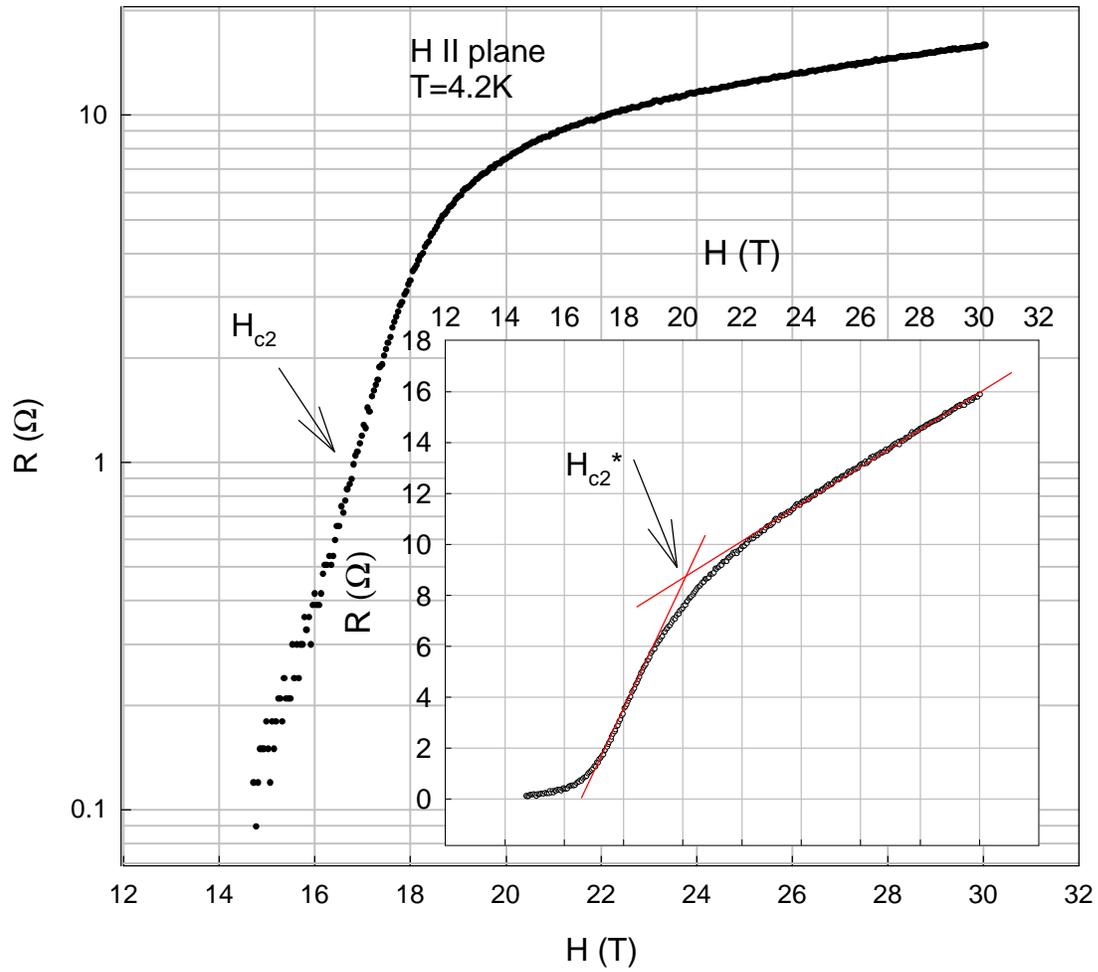
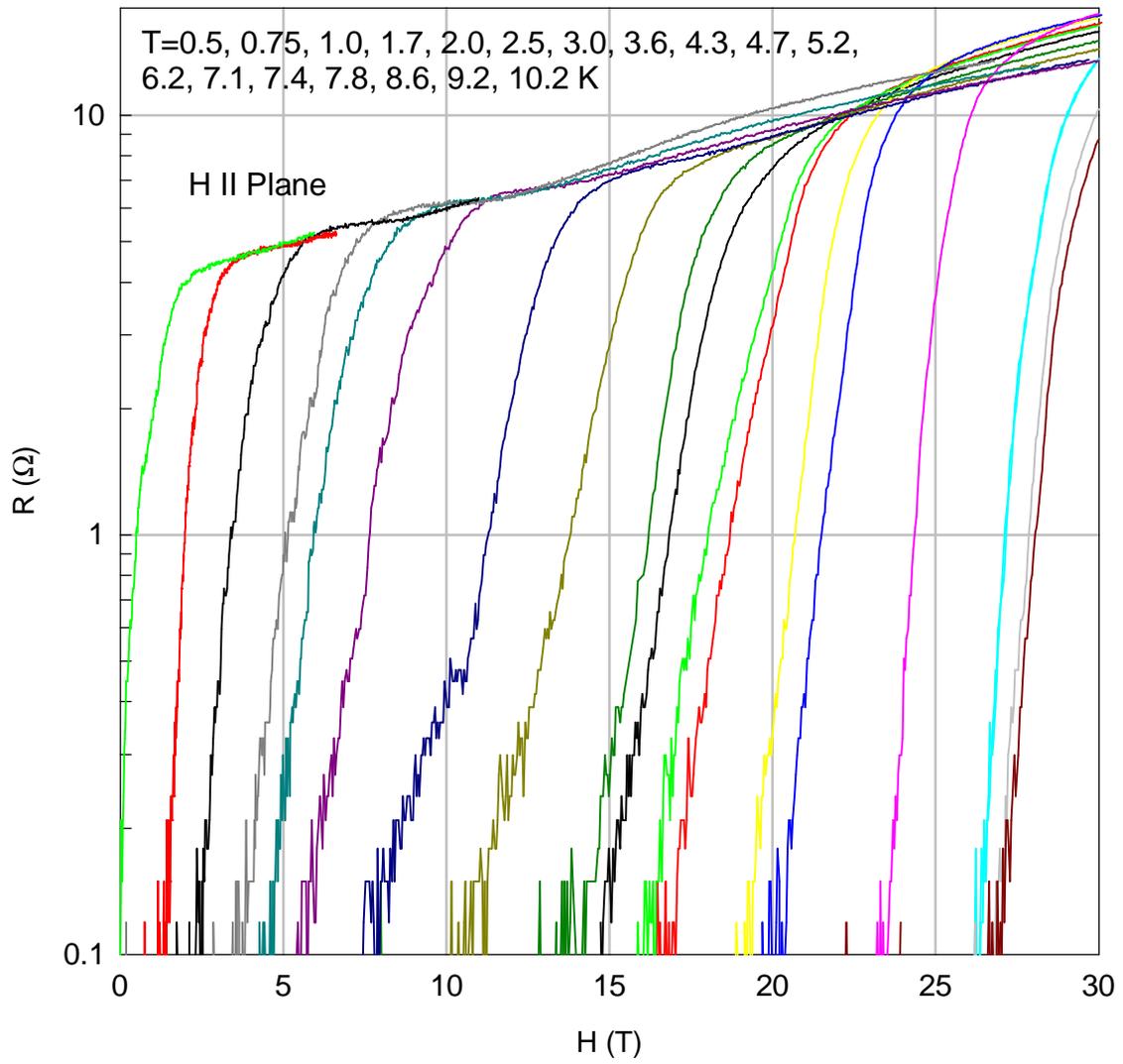
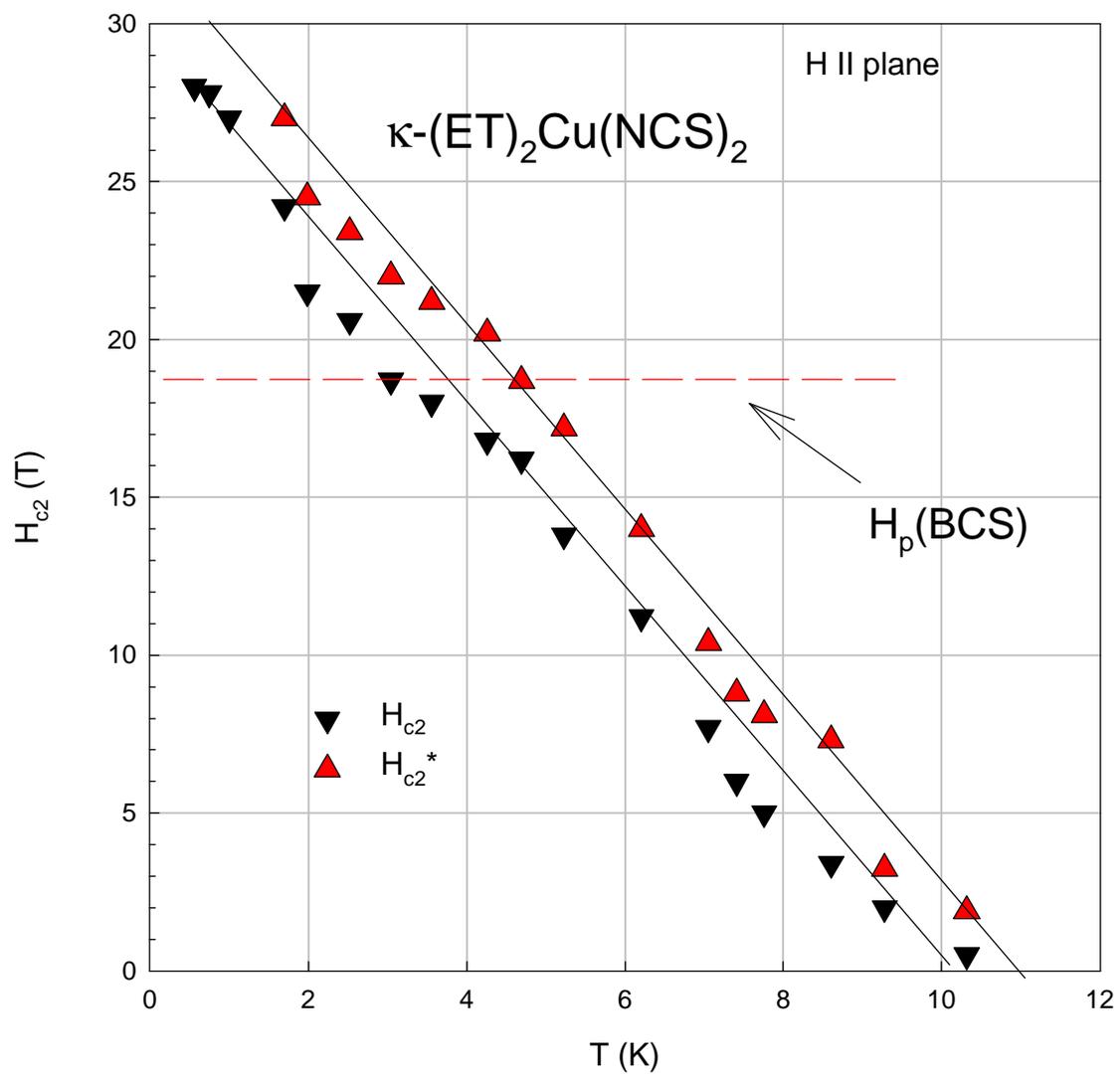
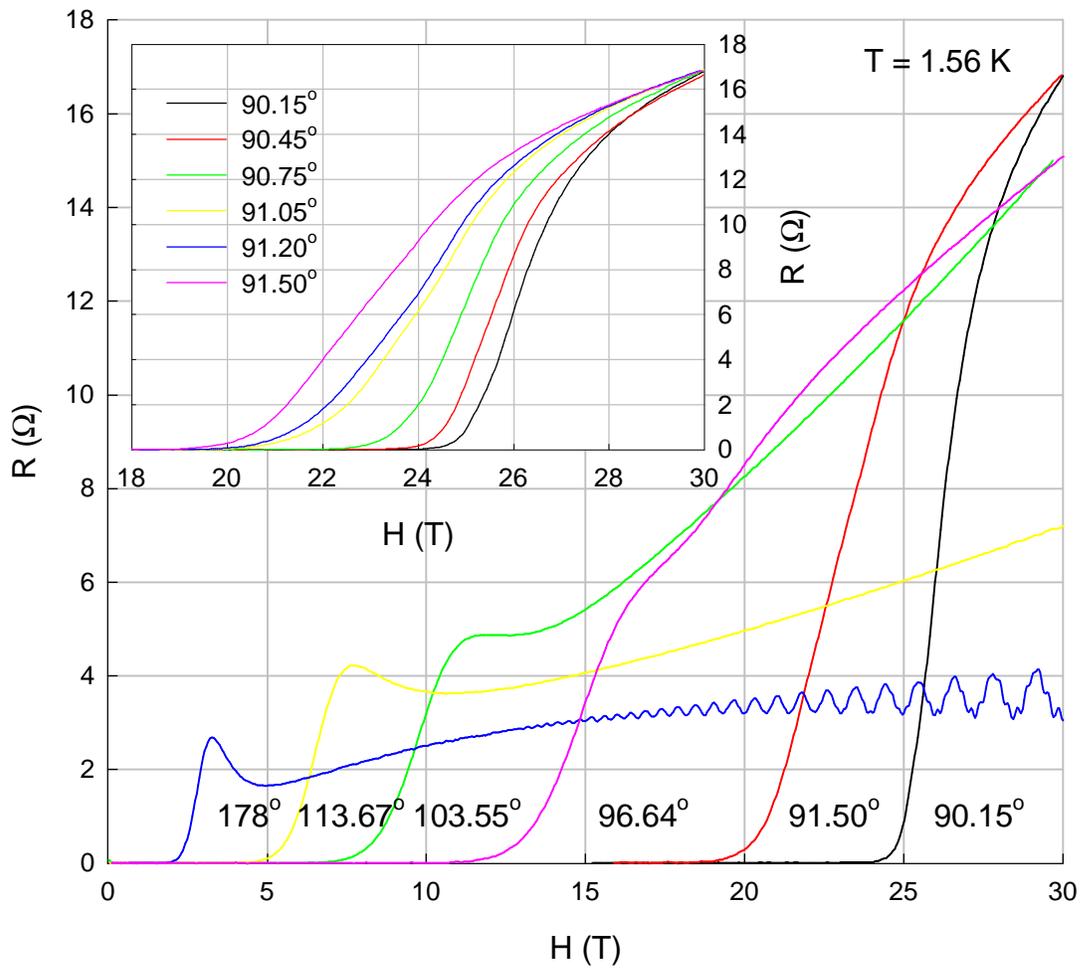
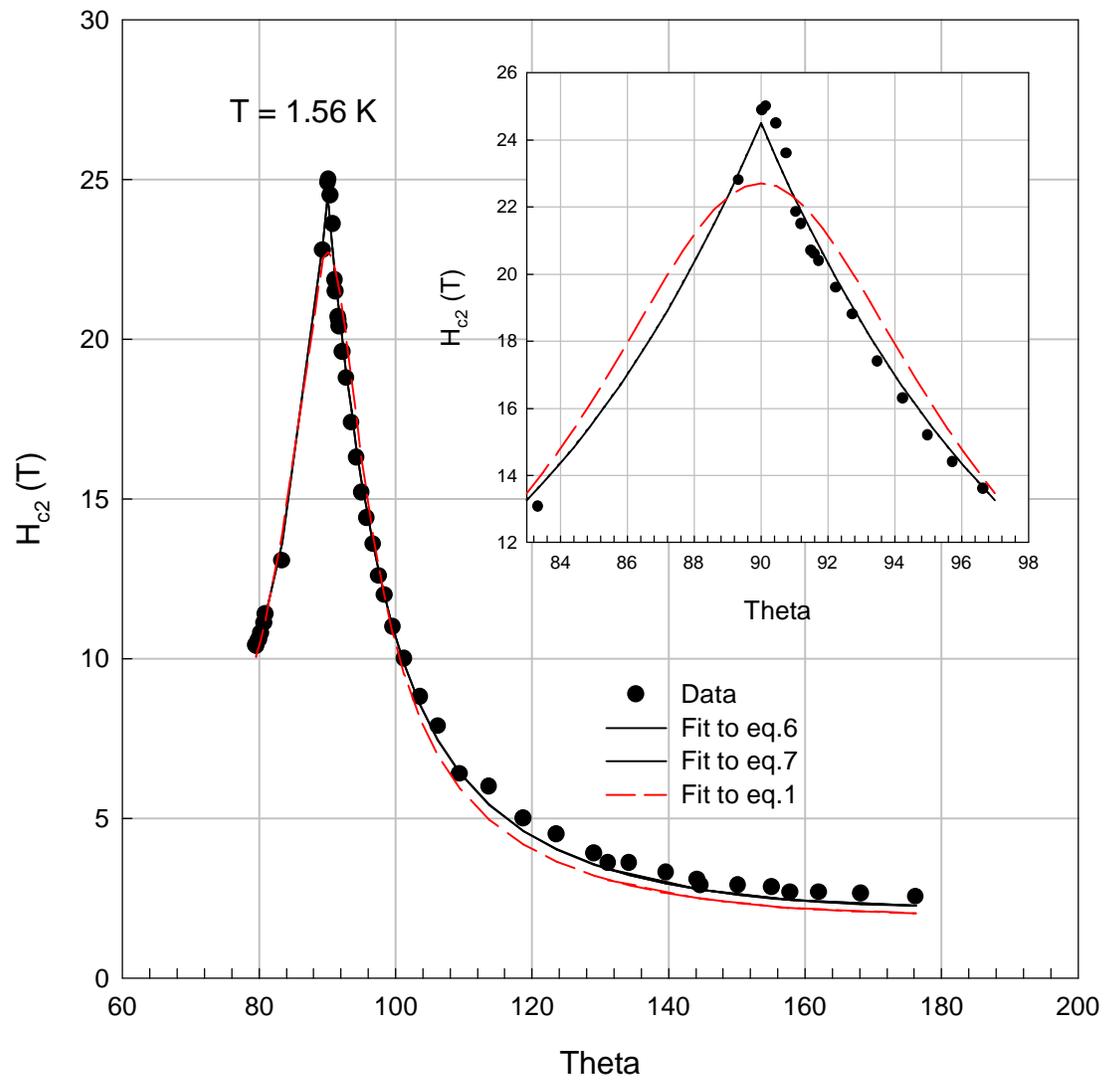
\end{document}